\documentclass[10pt,twocolumn]{article} 
\usepackage{simpleConference}
\usepackage{times}
\usepackage{graphicx}
\usepackage{amssymb}
\usepackage{url,hyperref}
\usepackage{bookmark}
\usepackage[font=small,labelfont=bf]{caption}
\usepackage{comment}
\usepackage{authblk}

\begin{document}

\graphicspath{ {figures/} }

\title{A knowledge graph representation learning approach to predict novel kinase-substrate interactions}

\author[1]{Sachin Gavali A\thanks{saching@udel.edu}}
\author[2]{Karen Ross A\thanks{ker25@georgetown.edu}}
\author[1]{Chuming Chen B\thanks{chenc@udel.edu}}
\author[1]{Julie Cowart}
\author[1]{Cathy H. Wu}
\affil[1]{Center for Bioinformatics and Computational Biology, University of Delaware, Newark, DE}
\affil[2]{Georgetown University Medical Center, Washington DC}

\renewcommand\Authands{ and }

\maketitle
\thispagestyle{empty}

\begin{abstract}

  The human proteome contains a vast network of interacting kinases and substrates. Even though some kinases have proven to be immensely useful as therapeutic targets, a majority are still understudied. In this work, we present a novel knowledge graph representation learning approach to predict novel interaction partners for understudied kinases. Our approach uses a phosphoproteomic knowledge graph constructed by integrating data from iPTMnet, Protein Ontology, Gene Ontology and BioKG. The representation of kinases and substrates in this knowledge graph are learned by performing directed random walks on triples coupled with a modified SkipGram or CBOW model. These representations are then used as an input to a supervised classification model to predict novel interactions for understudied kinases. We also present a post-predictive analysis of the predicted interactions and an ablation study of the phosphoproteomic knowledge graph to gain an insight into the biology of the understudied kinases.

\end{abstract}

\section{Introduction}
Proteins are a fundamental building block of the complex molecular machinery employed by all living organisms. The collection of all the possible proteins that can be synthesized by an organism is known as the proteome \cite{proteome}. Proteins interact with each other through distinct biochemical events to actuate the desired biological functions. Protein post-translational modification (PTM) is one such biochemical event that has played a major role in almost all the biological functions \cite{Uversky_2013}.

Fundamentally, any given PTM event is made up of two members - an enzyme and a substrate. An enzyme is a protein
responsible for facilitating the PTM event and the substrate is the protein undergoing the post-translational
modification. Among all the types of PTM events, phosphorylation is the most common and well-studied and is implicated
in a majority of cellular functions [3]. Phosphorylation is carried out by a class of enzymes known as kinases.
Previously it was believed that the kinase-substrate interaction networks are fairly linear, and perturbation of a
kinase would primarily affect its immediate substrate. But recent studies have shown that these interaction networks are
highly interconnected and perturbation of a particular kinase or a substrate has the potential to affect large parts of
the network [4].

With the advent of techniques such as mass spectrometry based high throughput proteomics, many new phosphorylation sites
have been identified [5] but identifying kinases that phosphorylate these sites remains a challenging problem.
Experimental studies on kinase-substrate interactions are time-consuming and expensive and most research has been
focused on a small subset of the ~550 protein kinases found in humans. Computational approaches that can accurately
predict novel kinase-substrate interactions have the potential to increase our understanding of the human proteome. This
increased understanding will in turn help accelerate identification of new therapeutic targets and the development of
accompanying drugs to modulate these targets.

To this date, many tools have been developed to predict kinase-substrate interactions. Tools such as Scansite [6],
NetPhospK [7], PPSP [8], GPS [9, 10] and PredPhosph [11] rely on the properties of protein sequences around the
phosphorylation site also known as ”sequence motifs”, to predict kinases most likely to be associated with the given
phosphorylation site. But kinase-substrate interactions involve much more than sequence motifs and hence it is necessary
to include contextual factors when making these predictions. Thus tools such as NetworKIN [12], PhosphoPICK [13],
PhosphoPredict [14] and HeteSim [15] were developed that combine sequence and contextual information to make better
predictions. But many of the above tools have significant limitations in terms of kinome coverage. This is partly due to
the fact that these tools primarily rely on properties that can only be directly mapped to the kinases or substrates.
Understudied kinases by their very nature have limited information and hence are not annotated with these properties
making it difficult to use these tools.

Inspired by recent advancement in deep learning a new generation of tools are being developed to address these
shortcomings. DeepKinZero [16] is a tool that takes inspiration from deep learning techniques in computer vision and
employs a zero shot learning approach to transfer knowledge from well known kinases to understudied kinases. But
similar to the first generation tools, it relies primarily on sequence information. LinkPhinder [17] takes a
significantly different approach and formulates the task of predicting kinase-substrate interactions as a
link-prediction task. It considers kinases, substrates and phosphorylation sites to be constituting a knowledge graph
and uses knowledge graph completion algorithms to predict possible kinase-substrate interactions. A significant
limitation of all the above tools is that they do not take advantage of the long range dependencies between kinases and
substrates that are encoded in existing kinase-substrate interaction networks. In addition to this, they also fail to
model the deeper biological connections that are only evident by looking at the vast body of biomedical knowledge being
collected and organized in semantic databases such as Gene Ontology and Protein Ontology. 

In recent years, there has been a significant increase in the amount of biological data. This has made it increasingly
difficult to organize and derive knowledge from this data. Subsequently, semantic technologies that define a set of
standards for organizing and linking data were adopted. Using such linked (semantic) data can provide us with knowledge
that cannot be derived purely from protein sequences. They can help us craft algorithms that can truly capture the
biological roles of kinases and substrates. In this work we present a novel approach of learning from semantic data. 
Since the goal of this work was to investigate if knowledge graph/semantic data can be useful in predicting
kinase-substrate interactions we simplified the task by only predicting interactions at the kinase/substrate level
instead of the kinase/phosphorylation site level. Nevertheless, we think that the kinase/substrate representations
learned by our approach can be combined with tools working at sequence level such as DeepKinZero to obtain better
predictions at finer resolutions.

\section{Methods}

\subsection{Data}

We construct the knowledge graph by including data from iPTMnet \cite{iptmnet}, Protein Ontology (PRO)
\cite{pro_ontology}, Gene Ontology (GO) \cite{gene_ontology} and BioKG \cite{biokg}. To begin with, we use human PTM
data [Taxon code - 9606] from iPTMnet to construct a kinase-substrate interaction network. The iPTMnet data contains
26411 phosphorylation PTM events. Any given kinase-substrate pair can have multiple PTM events. We normalize these
events to triples in the form of \textit{kinase \textrightarrow \  phosphorylates \textrightarrow \ substrate}. 

PRO defines protein classes and represents the hierarchical relationships among proteins, protein forms (proteoforms)
and protein complexes within and across species (PMID:28150233). Thus using PRO data we construct triples in the form of
\textit{kinase/substrate \textrightarrow \ is\_a \textrightarrow \  pro\_entity} and inverse triples in the form of
\textit{pro\_entity \textrightarrow \  has\_a \textrightarrow \ kinase/substrate} to capture evolutionary relationships
among the proteins in our knowledge graph.

Gene Ontology organizes biological knowledge by specifying a controlled vocabulary to precisely describe the biological
processes, molecular functions and subcellular localizations associated with gene products. Using GO we create triples
in the form of \textit{kinase/substrate \textrightarrow \ annotated\_with \textrightarrow \  go\_term}. Since GO terms
themselves are arranged in the form of a directed acyclic graph (DAG), we create new triples in the form
\textit{go\_term\_a \textrightarrow \ is\_a \textrightarrow \  go\_term\_b} extending uptil the root of the GO tree to
capture the knowledge defined by the relational heirarchy of GO.

Similar to the above-mentioned data sources, there are many more data sources that can be integrated in our knowledge
graph. Rather than performing this integration ourselves, we decided to take advantage of BioKG which provides a
framework to automatically perform the data integration. Since BioKG is geared towards drug discovery analysis we
integrate only a subset of it. Specifically we include triples with following relations - \textit{protein-pathway
associations, protein-disease associations, protein-genetic disorder associations, disease-genetic disorder
associations, disease-pathway associations, protein-complex assocations and complex-pathway associations}.

Once the above knowledge graph is built, we use it as a data source to train a machine learning model to predict
interactions for understudied kinases. 

\subsection{Data preparation}

As mentioned in the previous section, we start with a kinase-substrate interaction network constructed using PTM data
and then enrich it with auxillary data to construct our knowledge graph. When training a machine learning model it is
neccessary to ensure proper seperation of training, validation and testing data to prevent information leakage. Thus,
even before we enrich the vanilla kinase-substrate network with auxillary data, we split the network into three
subnetworks - training, validation and testing. Training network contains "\textit{kinase \textrightarrow \
phosphorylates \textrightarrow \ substrate}" triples in addtion to the triples from auxillary data. Validation and
testing networks contain only the kinase-substrate interaction triples in the form \textit{kinase \textrightarrow \
phosphorylates \textrightarrow \ substrate}. 

\subsection{Knowledge graph learning approach}
In recent years, many approaches to learn from knowledge graph have been proposed. These approaches can be broadly grouped into four categories - 1) Tensor decomposition, 2) Geometric distance, 3) Deep learning and 4) Random Walk \cite{kg_embedding_lp}. Tensor decomposition based approaches represent the entities and the relations as a giant 3D adjacency matrix (Tensor). This matrix is then decomposed into low dimensional vectors while still retaining the latent information about the graph structure and connectivity \cite{tensor_decomposition_1,tensor_decompisition_2}. Geometric distance based approaches learn an embedding of the knowledge graph by represent the relation between the head and tail as a geometric transformation in the latent space \cite{transe}. Deep learning based approaches represent the entities and relations using a low dimensional embedding vector. Instead of deriving these embeddings using tensor decomposition or geometric factorization, these models use a neural network to optimize the embeddings to predict the probability of a triple in the knowledge graph being true or false \cite{conve}. Random walk based approaches take inspiration from advancements in natural language processing. They involve sampling a series of nodes (entities) from the knowledge graph. These series of nodes can be thought of as sentences in a language with every node representing a word in the sentence. These sentences are then used as an input corpus for a language model such as word2vec \cite{word2vec} to learn a dense embedding for every node in the graph \cite{deepwalk}.

A glaring short-coming of the random walk based approaches is that they do not take into account the triple structure of the knowledge graph. Specifically, existing methods such as DeepWalk \cite{deepwalk} and Node2Vec \cite{node2vec} do not consider the directionality and the heterogeneity encoded by a triple when performing the random walk. They treat the relations in a knowledge graph as any other node in the graph. Hence, they cannot adequately capture the semantic meaning of the entities in the knowledge graph. To alay these shortcomings, alternative approaches that rely on metapaths have been proposed \cite{metapath2vec}. But contrary to the simpler approaches such as DeepWalk and Node2Vec, the performance of metapath based approaches is highly dependent on the choice of metapath. Additionally, choosing a metapath requires an in-depth knowledge of the schema of the knowledge graph under study, further diminishing their utility.

\begin{figure}[htbp]
  \centerline{\includegraphics[width=0.30\textwidth]{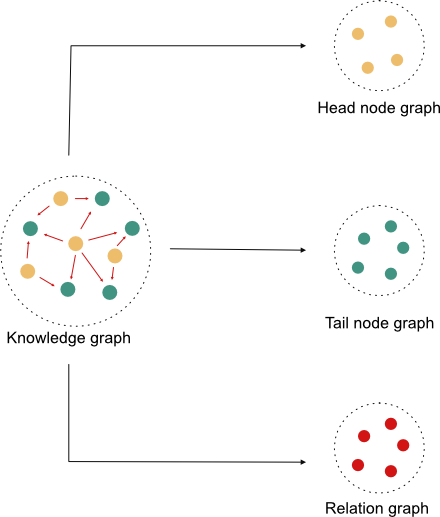}}
  \caption{Knowledge graph represented as an overlapping graph of heads, relations and tails} 
  \label{fig:kg_super_imposed}
\end{figure}

Hence, in this work, we propose a modified random walk based approach that takes inspiration from DeepWalk \cite{deepwalk}, one of the simplest knowledge graph learning algorithms, to learn a representation of kinases and substrates in our phosphoproteomic knowledge graph. The fundamental assumption of any random walk based approach is that entities with similar meaning occur in similar contexts. But in a knowledge graph, the context is not only defined by the connectivity, but also by the type and the direction of the relationships. Hence, our approach makes a slightly different assumption. It assumes that the heterogeneous knowledge graph is a superimposition of three distinct graphs. The first graph contains only head entities, the second graph contains only relations and the third graph contains only tail entities. [Figure \ref{fig:kg_super_imposed}]. The heterogeneous knowledge graph can then be thought of as a function of the latent interactions between the entities from each of these three sub-graphs. To model this function, we modify the manner in which random walks are performed. Instead of sampling a series of nodes using traditional random walks, we sample a series of ``triples'' by performing a triple walk. This series of triples is then used as an input to a modified skip gram model to learn an embedding of all the entities and relations in the knowledge graph.

\subsection{Deepwalk overview} \label{deepwalkapp}
Since our approach is inspired by DeepWalk approach, it is essential to understand all the steps that constitute the DeepWalk algorithm. On a very high level, the DeepWalk algorithm combines random walks on a graph with a language model such as Word2Vec \cite{word2vec} to learn a vector representation of every node in the graph. Since the Word2Vec model plays a major role in the DeepWalk algorithm, it is essential to understand the steps involved in training a Word2vec model.

\begin{figure}[htbp]
  \centerline{\includegraphics[width=0.50\textwidth]{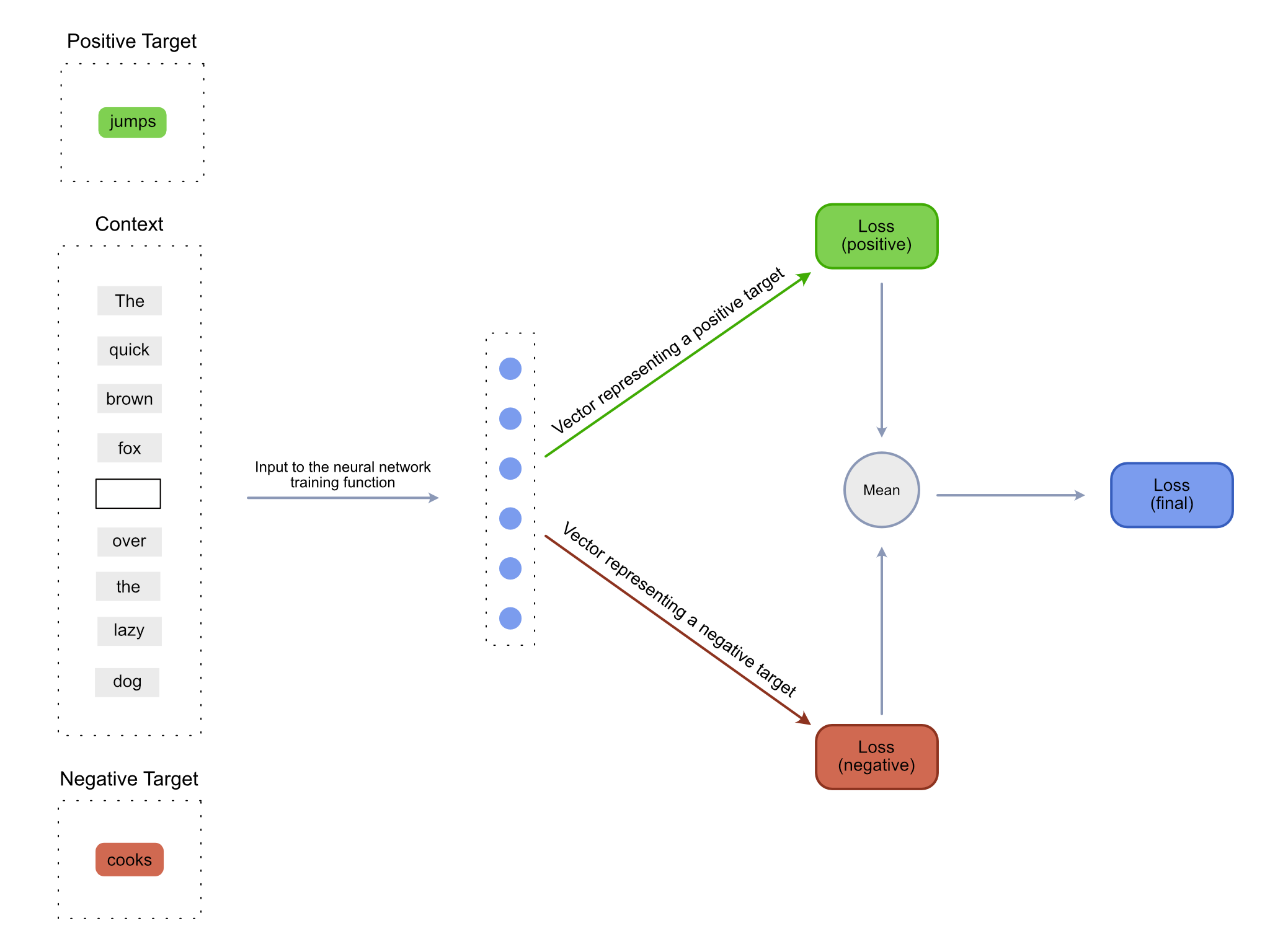}}
  \caption{Word2vec model (CBOW)} 
  \label{fig:word2vec_cbow}
\end{figure}

Word2Vec is a simple model used to learn dense vector embeddings of words \cite{word_embeddings} in a given language. At it's core it contains a single layered neural network that predicts if a particular word would occur in a given sentence. This task is similar to the task of filling the blanks in an incomplete sentence. For example, given an incomplete sentence - \textit{The quick brown fox \_\_\_\_\_ over the lazy dog}, the word2vec model tries to predict a word that would occur in the blank space [Figure \ref{fig:word2vec_cbow}]. The word to be predicted is known as the \textit{target} word and the words already present in the sentence are known as the \textit{context} words. The target word can be either a \textit{positive\_target} or a \textit{negative\_target}. A \textit{positive\_target} is a word that is definitely known to `'occur`' in the given blank space. A \textit{negative\_target} is a word that is definitely known to `'not occur`' in the given blank space. The \textit{negative\_target} is created by randomly choosing a word from all the words constituting the vocabulary of the language. The length of the sentence is known as the \textit{window\_size} or the \textit{context\_size} of the model and the number of words in the entire corpus is known as the \textit{vocabulary} of the language. 

So to recapitulate, the inputs to a function training the Word2Vec model are the \textit{context}, the \textit{positive\_target} and the \textit{negative\_target} [Figure \ref{fig:word2vec_cbow}]. This function then trains the model using the following three-step process.

\begin{enumerate}
  \item The \textit{context} words are used as an input to a single layered neural network to predict a vector representing the \textit{positive\_target}. The predicted vector is then compared with the vector of the ground truth positive word to calculate a \textit{positive\_score}. This score is then used to calculate a \textit{positive\_loss}.
  \item The same \textit{context} words, coupled with the same neural network are then used to predict a vector representing the \textit{negative\_target}. Then similar to step 1, the predicted vector is compared with the ground truth to calculate a \textit{negative\_score}. This score is then used to calculate a \textit{negative\_loss}. 
  \item The \textit{positive\_loss} and \textit{negative\_loss} are then combined using the \textit{mean} function to calculate the \textit{final\_loss}. This \textit{final\_loss} is then used to backpropogate the errors and adjust the weights and biases of the neural network as well the embedding vectors of the context words.
\end{enumerate}

The above process is repeated for every word in each sentence of the entire corpus. As the training progresses, the model learns which \textit{target} word occur in which \textit{context}. Once the model training is complete, the embeddings vectors of \textit{context} words are retrieved to be used as a part of further downstream analysis.

\begin{figure}[htbp]
  \centerline{\includegraphics[width=0.50\textwidth]{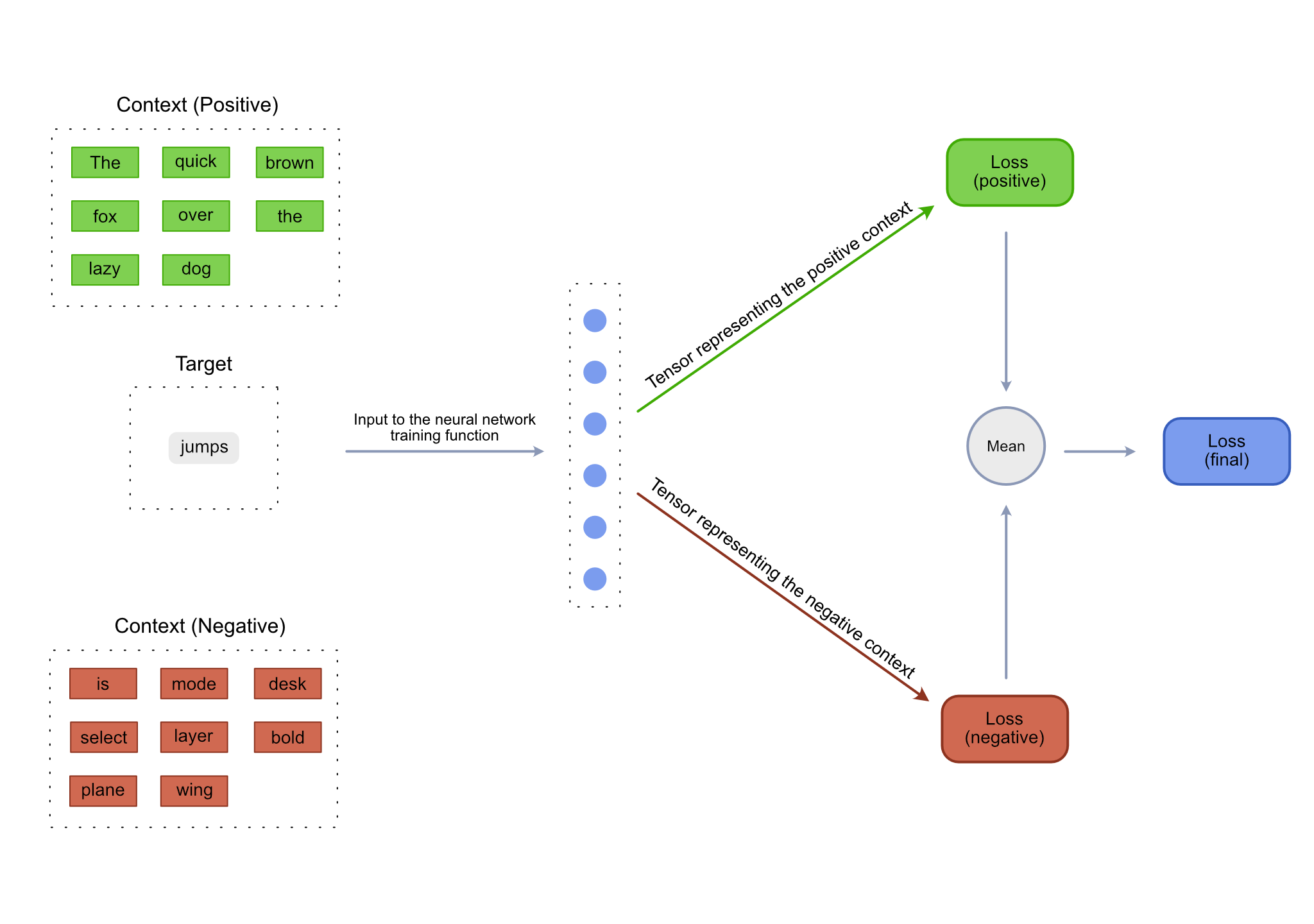}}
  \caption{Word2vec model (SkipGram)} 
  \label{fig:word2vec_skipgram}
\end{figure}

\begin{figure}[htbp]
  \centerline{\includegraphics[width=0.28\textwidth]{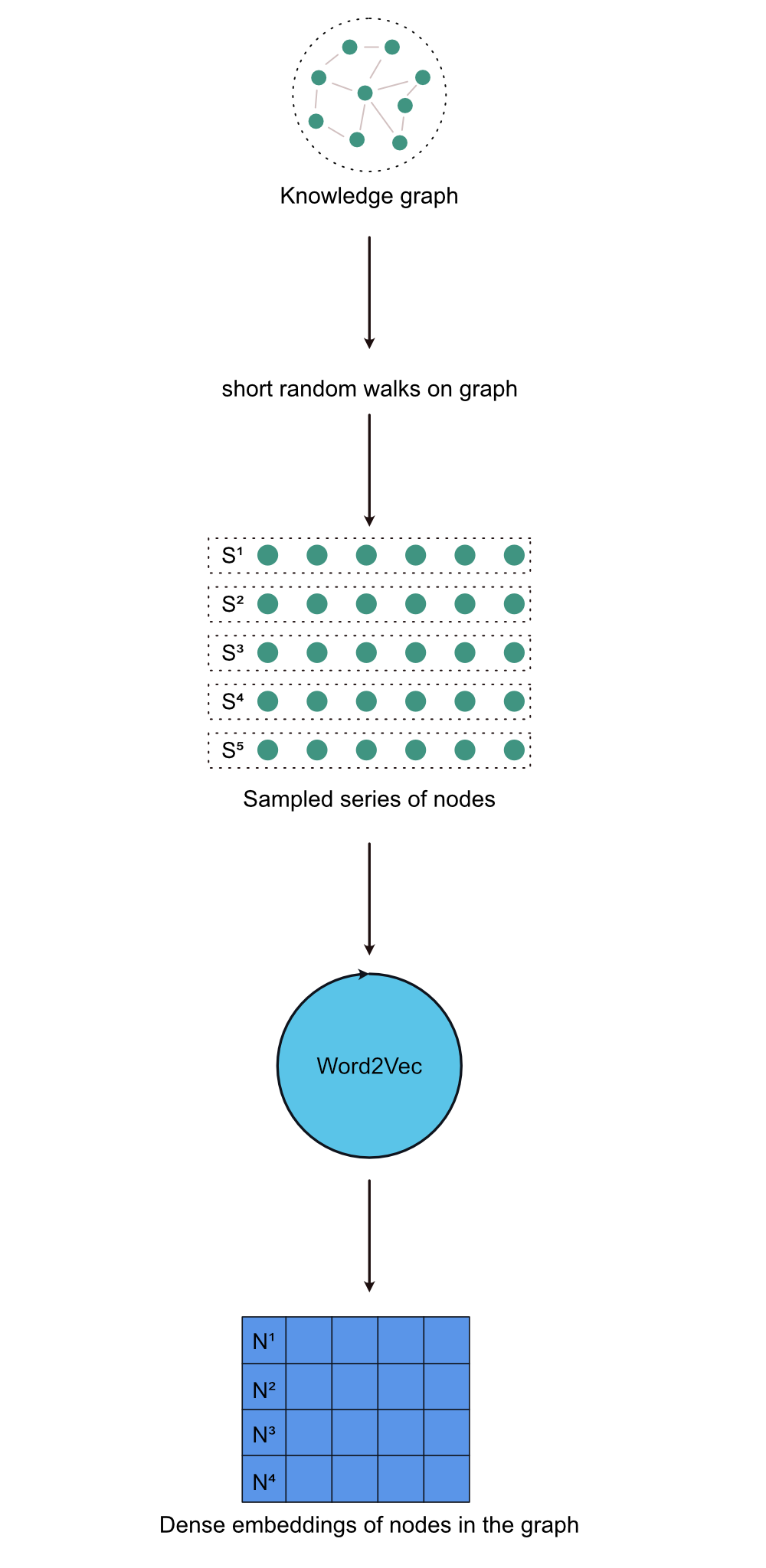}}
  \caption{DeepWalk model} 
  \label{fig:deepwalk}
\end{figure}

Word2Vec model has two variants, CBOW (Continuous Bag of Words) and SkipGram. The model described above is the CBOW variant of the Word2Vec model. SkipGram variant is the exact inverse of the CBOW variant. In the SkipGram variant, instead of predicting a \textit{target\_word}, the model predicts a \textit{target\_context}. So the inputs to the training function of the SkipGram variant are the \textit{target\_word}, the \textit{positive\_context} and the \textit{negative\_context} [Figure \ref{fig:word2vec_skipgram}]. The \textit{positive\_context} contains the words that are definitely known to `'occur`' around the \textit{target\_word} and \textit{negative\_context} contains the words that are definitely known to `'not occur`' around the \textit{target\_word}. The \textit{negative\_context} is created by randomly sampling words from the vocabulary of the language. The remaining steps in the training function are similar to the CBOW variant except for the inputs to the single layered neural network. Contrary to the CBOW variant, the input to the neural network is a \textit{target\_word} and the output is a tensor representing the \textit{context}. This tensor is then compared with the tensor of the ground truth positive and negative contexts to obtain a positive and negative score. These scores are then used to calculate the respective losses, which are combined to get the final loss. 

The authors of the DeepWalk algorithm hypothesized that the language models work by sampling from a hidden unobservable language graph \cite{deepwalk}. This means that every graph can be thought of as encoding the semantics of a hidden unobservable language. So, the first step of the DeepWalk algorithm is to perform short random walks on the graph to sample a series of nodes. The random walks performed in DeepWalk are a classical Markovian process \cite{markovian} i.e. the probability of selecting the next node in the walk is only dependent on the currently selected node. Now, these series of nodes can also be thought of as a series of words adding up to form a complete sentence. Thus performing \textit{N} random walks on the graph can be thought of as sampling a set of \textit{N} sentences from a graph. These sentences i.e. series of words are now used as an input to a language model such as the Word2Vec model to learn a dense vector representation of every node in the graph [Figure \ref{fig:deepwalk}].

\subsection{TripleWalk approach}
As described earlier, a unit of information in a knowledge graph is encoded by a triple in the form of \textit{head \textrightarrow \ relation \textrightarrow \ tail}. Thus, to learn an effective representation of a knowledge graph it is essential to consider this triple structure. The TripleWalk algorithm modifies the DeepWalk algorithm to effectively exploit this triple structure. It does so by modifying the process of performing the random walks and also the process of using these random walks to train the Word2Vec model.

In the DeepWalk approach of performing random walks, the probability of choosing the next node in the walk is only dependent on the current node. The type of node i.e. if it is a \textit{head} or \textit{tail} does not have any influence on the walking process. Further, the DeepWalk approach also does not consider the directionality of the edges. For example, at any point of time when the walker is on a \textit{head} or a \textit{tail} node, it has a choice of either selecting one of the \textit{tail} nodes that come after a \textit{head} node or one of the \textit{head} nodes that come before the \textit{tail} node. Since the walker does not take into account the directionality of the edges, it has an equal probability of choosing a \textit{head} node or a \textit{tail} node. If it samples a \textit{head} node then it inadvertently ends up breaking the semantic organization of the underlying graph. 

Contrary to the DeepWalk approach, the TripleWalk approach does not sample one node at a time, but samples one triple at a time. Thus, the probability of choosing the next triple in the walk is dependent only on the currently selected triple. Further, the TripleWalk approach also considers the directionality of the relation between triples. At any point of time, given a triple sequence $T_1$ \textrightarrow $T_2$ \textrightarrow $T_3$ \textrightarrow $T_4$ \textrightarrow $T_5$, a triple walker at position $T_3$ will only sample $T_4$ and not $T_2$. Thus, by sampling one triple at a time and by considering the directionality of the triple relations, the TripleWalk approach is able to preserve the semantic structure of the underlying graph when sampling a sequence to be used in the Word2Vec model.

\begin{figure}[htbp]
  \centerline{\includegraphics[width=0.48\textwidth]{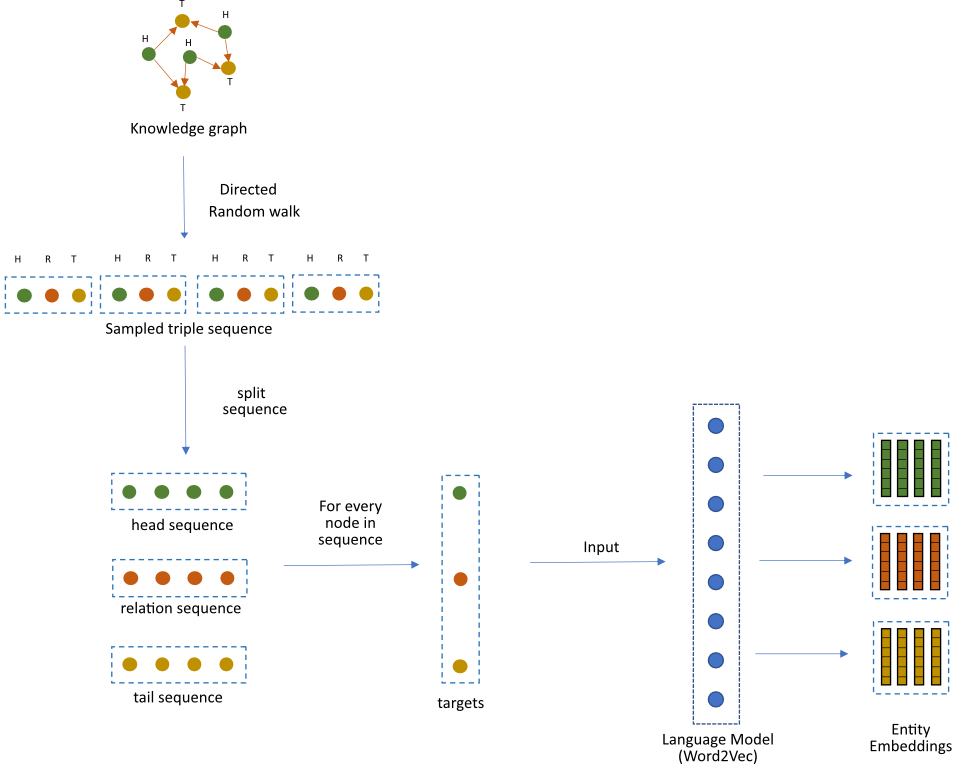}}
  \caption{TripleWalk model} 
  \label{fig:triplewalk}
\end{figure}

Once these triple are sampled, the next step is to learn an embedding of entities that make up these triples. For this we lean on our assumption mentioned earlier, that considers a knowledge graph as a combination of three distinct graphs - \textit{head\_graph}, \textit{relation\_graph} and \textit{tail\_graph} that hold the \textit{heads}, \textit{relations} and \textit{tails} respectively [\ref{fig:kg_super_imposed}]. To model this assumption, the sampled triple sequence \textit{$S = \{(h_1,r_1,t_1), (h_2,r_2,t_2), (h_3,r_3,t_3), (h_4,r_4,t_4)\}$} is split into three independent sequences holding heads (\textit{$H = \{(h_1,h_2,h_3,h_4)\}$}), relations (\textit{$R = \{(r_1,r_2,r_3,r_4)\}$}) and tails (\textit{$T = \{(t_1,t_2,t_3,t_4)\}$}) respectively. These three independent sequences are then used to train a modified Word2vec model [\ref{fig:triplewalk}].

\begin{figure}[htbp]
  \centerline{\includegraphics[width=0.45\textwidth]{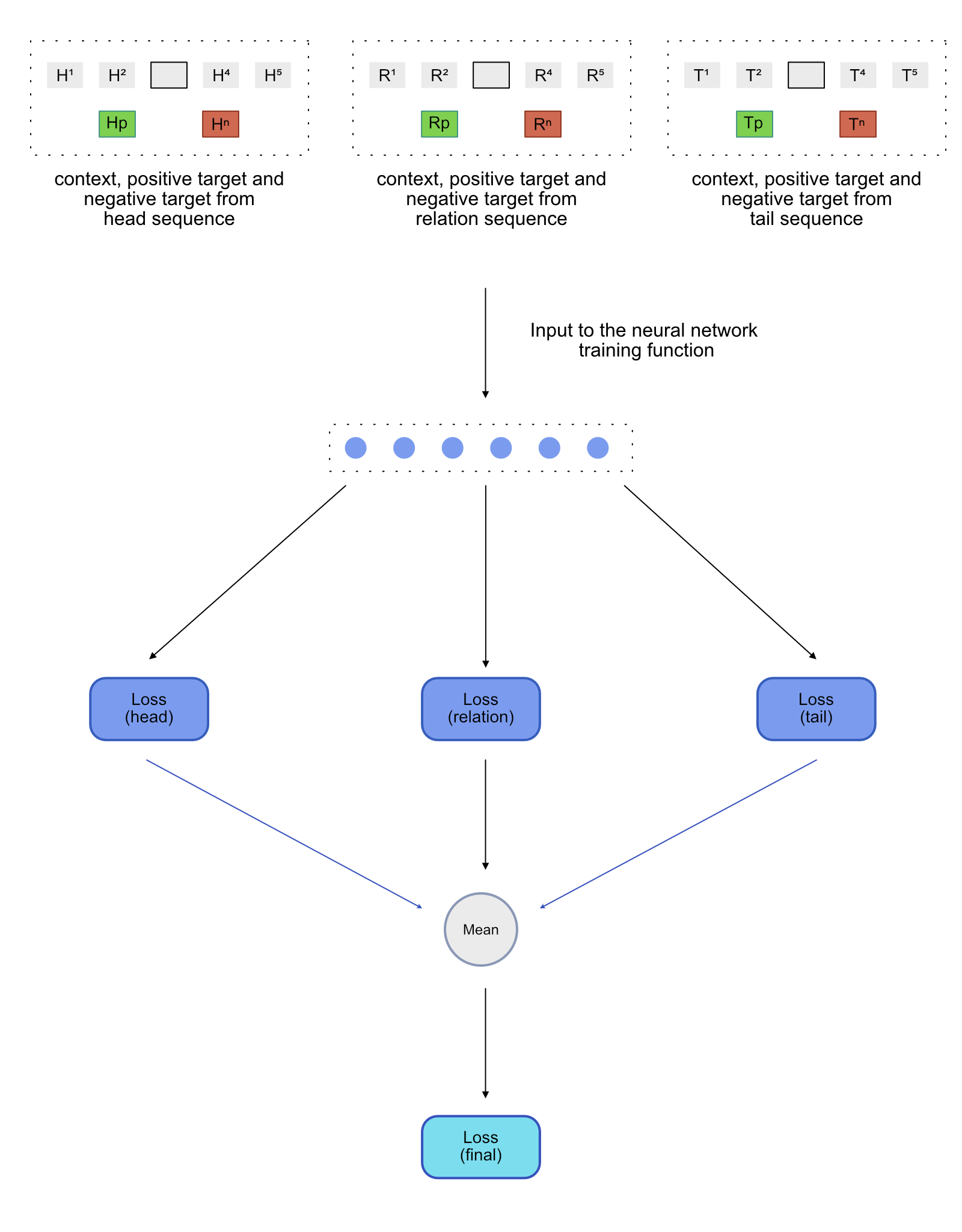}}
  \caption{Modified Word2Vec model} 
  \label{fig:modified-word2vec}
\end{figure}

As described in section \ref{deepwalkapp}, the input to the function used to train Word2Vec model are the \textit{context}, the \textit{positive\_target} and the \textit{negative\_target}. Thus, for every independent sequence, a \textit{context}, a \textit{positive\_target} and a \textit{negative\_target} is created. This gives us a set of three contexts - \textit{head\_context} ($H_c$), \textit{relation\_context} ($R_c$) and \textit{tail\_context} ($T_c$), a set of three positive targets - \textit{head\_pos\_target} ($H_p$), \textit{rel\_pos\_target} ($R_p$) and \textit{tail\_pos\_target} ($T_p$) and a set of three negative targets - \textit{head\_neg\_target} ($H_n$), \textit{rel\_neg\_target} ($R_n$) and \textit{tail\_neg\_target} ($T_n$). All the above contexts and targets are then used as an input to a function used to train the Word2Vec model [Figure \ref{fig:modified-word2vec}]. 

The function used to train the Word2Vec model is similar to the one used in DeepWalk model. As described earlier, the DeepWalk training function optimizes the embedding vectors of the context nodes using a single layered neural network. Similar to the DeepWalk training function, the TripleWalk training function also contains a single layered neural network, but instead of optimizing a single context embedding, it jointly optimizes the three independent sets of context embeddings corresponding to the head, relation and tail contexts. To do so, it follows a four-step process 

\begin{enumerate}
  \item The \textit{head\_context} ($H_c$) along with the \textit{head\_pos\_target} ($H_p$) and \textit{head\_neg\_target} ($H_n$) are used to calculate a \textit{head\_loss}.
  \item The \textit{relation\_context} ($R_c$) along with the \textit{rel\_pos\_target} and \textit{rel\_neg\_target} ($R_n$) are used to calculate a \textit{realtion\_loss}.
  \item The \textit{tail\_context} ($T_c$) along with the \textit{tail\_pos\_target} ($T_p$) and \textit{tail\_neg\_target} ($T_n$)are used to calculate a \textit{tail\_loss}.
  \item All these losses are then combined using the \textit{mean} function to obtain a \textit{final\_loss}. This \textit{final\_loss} is then used to backpropogate the errors and adjust the weights and biases of the neural network as well the embedding vectors of all the three contexts.
\end{enumerate}

The above process is repeated for every triple sequence sampled by the TripleWalk algorithm to minimize the \textit{final\_loss}. Once the training process is complete, the embedding vectors of the head context, relation context or the tail context are retrieved to perform further downstream analysis.

\subsection{Supervised learning}
The task of identifying new interactions in the kinase-substrate interaction network can be generalized to a binary classification task of predicting if a given interaction is true or false. For this, we use a classical supervised machine learning algorithm - Random Forest \cite{rf}. The input to the model is an embedding vector representing the target interaction and the output is a binary value representing the plausibility of the interaction being true or false. To construct an embedding vector denoting this interaction, we retrieved a list of all the \textit{kinase \textrightarrow \ phosphorylates \textrightarrow \ substrate} triples from the knowledge graph. Then, in a given triple we retrieve the embedding vector for \textit{kinase} entity from the \textit{head\_context} embeddings and for the \textit{substrate} entity from the \textit{tail\_context} embeddings. Then according to the approach described by the authors of Node2Vec algorithm \cite{node2vec}, we combine these embeddings using the \textit{hadamard} ($\odot$) operator to obtain the final interaction embedding vector (${\vec{I}}$).

\subsection{Negative sampling}
Since the supervised model is a binary classification model, we also need to have negative samples to represent the interactions that have a lower likelihood of being true. But adequate ground truth data about negative interactions is not available. Hence, it is important to adopt a well-thought-out approach to generating negative samples. Since knowledge graphs contain only positive samples, some approaches to generating negative samples have been proposed \cite{negative_sampling}. The most simple approach being corrupting a triple by randomly changing the head, relation or tail. 

\begin{figure}[htbp]
  \centerline{\includegraphics[width=0.50\textwidth]{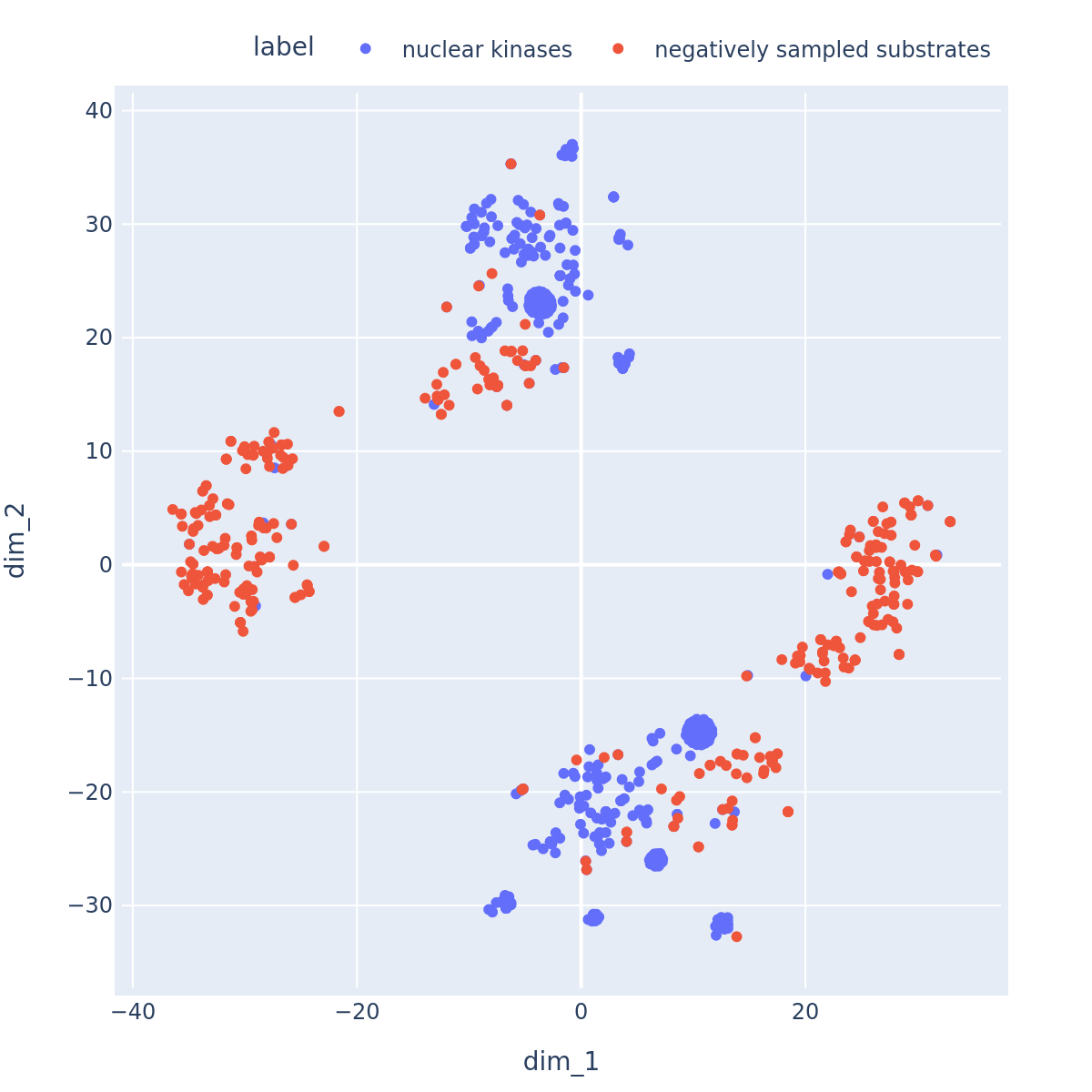}}
  \caption{tSNE plot to visualize the embeddings of kinases and their analogous sampled substrates using the negative sampling technique described in section 2.7} 
  \label{fig:negative_sampling}
\end{figure}

In this case, we were only interested in negative samples involving kinase-substrate interaction triples. Initially, we trained the supervised classification model using the naive negative sampling approach of corrupting the tails. The resulting model did not perform sufficiently (AU-ROC = 0.57). We hypothesized that the drop in performance was due to the fact that randomly corrupting the tails of triples did not yield samples that truly represented the underlying biology of a kinase-substrate interaction. Also, since the number of unknown kinase-substrate interactions is very high, there is an increased likelihood of true positive samples being labelled as negative samples.  Hence, we decided to develop a better approach to generating negative samples. We assumed that if a kinase and a substrate were physically apart by being located in two distinct cellular components then the probability of them interacting is lower than if they were located in the same cellular component. To model this assumption, we generated negative samples using the following four-step process:

\begin{enumerate}
    \item Create a filtered knowledge graph containing only kinase-substrate interaction triples and triples
    from the cellular component subtree of the GO ontology.

    \item Generate an embedding of every kinase and substrate in terms of its subcellular location by performing
    graph representation learning on this knowledge graph.

    \item Using this embedding, for every kinase sample N substrates that are as far away as possible in the
    embedding space by use cosine similarity to calculate the distance between a kinase and a substrate.

    \item To balance out the possibility of the model being biased towards the subcellular location, combine the
    above sampled list with ground truth negative samples from negatome - a database containing manually curated
    negative samples \cite{negatome}.

    \item Finally, sample from the above list to create a definitive list of negative interactions.

\end{enumerate}

After generating the negative samples using the above approach, we needed to verify if the generated negative samples contained substrates in cellular compartments that where distinct from kinases. Hence, we created a list of kinases that where located in the nucleus of the cell. Then we retrieved the negative interaction partners (substrates) for these kinases. We then visualized the embedding vectors of these kinases and substrates using a tSNE plot [Figure \ref{fig:negative_sampling}].

\subsection{Model training and evaluation}

\begin{figure}[htbp]
  \centerline{\includegraphics[width=0.33\textwidth]{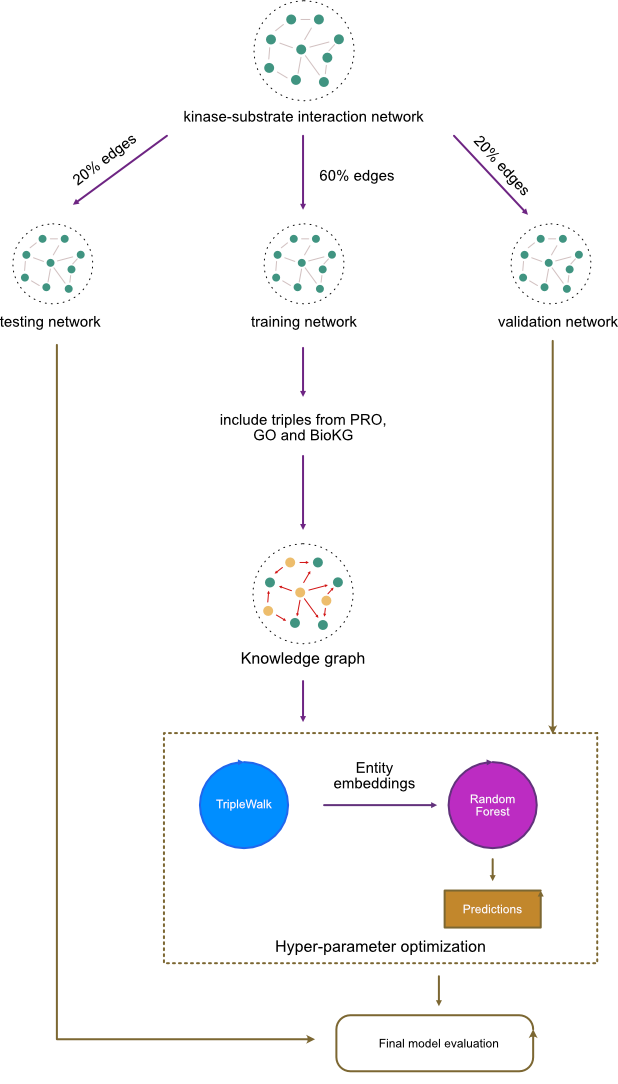}}
  \caption{TripleWalk model training and hyperparameter optimization} 
  \label{fig:triple_walk_training}
\end{figure}

Before we start training the model we split the kinase-substrate interaction data into training (60\%), validation (20\%) and testing (20\%) sets. We annotate only the training data with auxillary triples to construct the full knowledge graph. We then perform unsupervised learning to learn vector embeddings of kinases and substrates in this knowledge graph. These embeddings are then used to train a random forest model to identify the novel kinase-substrate interactions. We use the training and validation set to perform the hyperparameter tuning of both unsupervised and supervised component. We use the testing set to evaluate the final model using Area Under Receiver Operating Characteristic Curve (AU-ROC) [Figure \ref{fig:triple_walk_training}].

\subsection{Evaluation of unsupervised learning component}
In addition to evaluating the final model, we also compare the unsupervised learning component (TripleWalk algorithm) with existing unsupervised knowledge graph learning methods. We compare our approach with two of the most commonly used random walk based approaches - DeepWalk \cite{deepwalk} and Node2vec \cite{node2vec}, one tensor decomposition based approach - DistMult \cite{distmult} and one distance based approach - TransE \cite{transe}. In addition to comparison with above methods, we also compare the SkipGram and the Continuous Bag of Words (CBOW) variants \cite{word2vec} of all the random walk based methods. Finally, in addition to evaluating the interaction prediction performance, we also evaluate the embeddings on following tasks.

\begin{enumerate}
    \item Kinase - Substrate classification : We formulate a binary classification task to classify kinases and substrates based on the learned embeddings. The classification task is a balanced classification task where we sample one substrate for each of the 408 human kinases. We then use the embeddings obtained by unsupervised learning to train a classifier to classify the entities as either a kinases or substrates.

    \item Enzyme classification : All kinases can be generalized as enzymes. Enzymes are classified into six broad categories and numerous sub-categories based on the chemical reactions they catalyze. These enzyme categories are represented by an Enzyme Classification (EC) number. we use these EC numbers to create three sub-categories for kinases present in our kinase-substrate interaction network. We then use the embeddings of the kinases to formulate a one vs rest classification task to classify kinases into their respective categories.

\end{enumerate}

\section{Results}

\subsection{Model hyperparameters and performance}
We optimize the parameters for the unsupervised component by using the Adaptive Asynchronous Halving Algorithm (ASHA)
\cite{asha} and for the supervised component using Random Grid Search algorithm. The best performing hyperparameters for the TripleWalk model are provided in table \ref{table:hyperparams}. 

\begin{table}
  \centering
    \scalebox{0.80}{
      \begin{tabular}{| p{4cm} | p{4cm} |}
         \hline
         \multicolumn{2}{|c|}{\textbf{Unsupervised (TripleWalk and SGNS)}} \\
         \hline
         \textit{Parameter} & \textit{Value} \\
         \hline
         Batch Size & 128 \\
         \hline
         Learning Rate & 0.0004 \\
         \hline
         Embedding Dimension & 256 \\
         \hline
         Random walk length & 17 \\
         \hline
         Random walks per node & 6 \\
         \hline
         Early stopping delta (loss) & 0.1 \\
         \hline
         Early stopping patience (epochs) & 5 \\
         \hline
         Number of negative samples per positive sample & 2.0 \\
         \hline
        \multicolumn{2}{|c|}{\textbf{Supervised (Random Forest)}} \\
        \hline
         \textit{Parameter} & \textit{Value} \\
        \hline
        Number of Estimators & 420 \\
        \hline
        Max depth & 176 \\
        \hline
        Split Criterion & Entropy \\
        \hline
        Max features & 128 \\
        \hline
        Min samples for split & 5 \\
        \hline
        Min samples at leaf & 4 \\
        \hline
      \end{tabular}
    }
    \caption{{Best performing hyperparameters for the TripleWalk (SkipGram) model.}}
    \label{table:hyperparams}
  \end{table} 

\begin{table*}
  \centering
  \scalebox{0.60}{
    \begin{tabular}{| p{6.0cm} | p{4.2cm} | p{4.2cm} | p{4.2cm} | p{6.0cm} |}
       \hline
       \textit{Model} & \textit{Interaction Prediction} & \textit{Kinase classification} & \textit{Enzyme classification} & \textit{Model Type}\\
       \hline
       TripleWalk (Skip Gram) & \textbf{\textit{0.79} $(\pm0.01)$} & 0.67 $(\pm0.03)$ & 0.59 $(\pm0.02)$ & Directed Random Walk \\
       \hline
       TripleWalk (CBOW) & \underline{\textit{0.71}} $(\pm0.03)$ & \textbf{\textit{0.84}} $(\pm0.02)$ & \textit{\underline{0.68}} $(\pm0.03)$ & Directed Random Walk  \\
       \hline
       DeepWalk (Skip Gram) & 0.60 $(\pm0.01)$ & 0.57 $(\pm0.01)$ & 0.61 $(\pm0.02)$ & Undirected Random Walk  \\
       \hline
       DeepWalk (CBOW) & 0.69 $(\pm0.01)$ & 0.72 $(\pm0.02)$ & \textbf{\textit{0.72}} $(\pm0.03)$ & Undirected Random Walk  \\
       \hline
       Node2Vec (Skip Gram) & 0.62 $(\pm0.01)$ & 0.60 $(\pm0.04)$ & 0.62 $(\pm0.03)$ & Biased Random Walk  \\
       \hline
       Node2Vec (CBOW) & 0.71 $(\pm0.01)$ & 0.70 $(\pm0.03)$ & 0.67 $(\pm0.03)$ & Biased Random Walk  \\
       \hline
       Distmult & 0.63 $(\pm0.01)$ & \underline{\textit{0.78}} $(\pm0.02)$ & 0.67 $(\pm0.04)$ & Tensor decomposition \\
       \hline
       TransE & 0.61 $(\pm0.01)$ & 0.57 $(\pm0.01)$ & 0.57 $(\pm0.03)$ & Geometric distance  \\
       \hline
    \end{tabular}
  } \caption{{{Comparitive performance of unsupervised learning components on interaction prediction, kinase
  classification and enzyme classification}}}
  \label{table:unsupervised_results}
\end{table*}

The results from training all the models using the best hyperparameters are shown in table \ref{table:unsupervised_results}. It can be observed that for our primary task of predicting a kinase-substrate interaction, the TripleWalk algorithm coupled with the SkipGram model outperforms all other types of algorithms. For the kinase classification task the TripleWalk algorithm coupled with CBOW model outperforms all other algorithms and for the enzyme classification task DeepWalk model coupled with the CBOW model gives the best results. It is worth noting that CBOW model has a much more consistent performance compared to the SkipGram model irrespective of the type of random walks. This might be due to the fact that in the CBOW model contrary to the SkipGram model, we predict the target word given the context words. This might have a regularising effect on the model, preventing it from learning the noise in the data and thus leading to a much more stable performance. It is also interesting to observe that Distmult, a tensor decomposition based approach shows competitive performance with random walk based methods, despite being a much simpler algorithm. 

After comparing our model with existing Random Walk based methods, we also compared our model with other kinase-substrate prediction models. One thing to note is that compared to existing models, our model can only make predictions at the \textit{kinase-substrate} level instead of the \textit{kinase-substrate-site} level. The predictions from our model are in the form of a three column vector containing - \textit{[kinase, substrate, probability]} and the predictions from existing models are in the form of a four column vector containing - \textit{[kinase, substrate, site, probability]}. Hence, to make a fair comparision, we choose the best scoring site among all the predicted sites for a given kinase-substrate pair to be the probability of the kinase and substrate interacting with each other. Another caveat to this comparison is that many of the existing models do not publish the training and testing datasets as well the hyperparameters used to train the model. Hence, we used the final predictions datasets published by the authors of LinkPhinder \cite{linkphinder}. The results from our comparision are shown in table \ref{table:model-compare}. It can be observed that the knowledge graph based models such as TripleWalk and LinkPhinder show a significantly better performance compared to the sequence based models such as NetPhospK \cite{netphosphk} and Scansite \cite{scansite}. Further, it can be observed that the TripleWalk model shows competitive performance compared to the LinkPhinder model despite not including sequence based features in the knowledge graph. This might point to the fact that it would be beneficial to combine the knowledge graph construction approach proposed by LinkPhinder with the knowledge graph learning algorithm proposed by TripleWalk to further improve the performance.

\begin{table}
  \centering
    \scalebox{0.80}{
      \begin{tabular}{| p{3cm} | p{1.5cm} | p{2cm} | p{2cm} | }
         \hline
         \textit{Model} & \textit{AU-ROC} & \textit{Precision} & \textit{Recall} \\
         \hline
         TripleWalk & 0.76 & 0.62 & 0.88 \\
         \hline
         LinkPhinder & 0.75 & 0.60 & 0.76 \\
         \hline
         NetPhospK & 0.52 & 0.61 & 0.17 \\
         \hline
         Scansite & 0.53 & 0.60 & 0.17 \\
         \hline
         NetworKIN & 0.55 & 0.59 & 0.36 \\
         \hline
      \end{tabular}
    }
    \caption{{Comparision of TripleWalk with existing kinase-substrate interaction prediction models.}}
    \label{table:model-compare}
  \end{table}

\paragraph{Ablation Study}

After building our models, we wanted to understand the factors that contribute to the improved predicitve performance. Model interpretability can be achieved by either building simpler models that are intrinsicly explainable or by post predictive analysis of the trained models. Since our framework consists of multiple models working together, the simplest approach to achieving interpretability in our system would be to quantify the change in predictive performance on changing the input data. For this we follow a two-part approach. In the first part we remove only a particular set of triples while keeping all other triples in knowledge graph. We then train the model using the hyperparameters shown in Table \ref{table:hyperparams}. In the second part, we keep triples related to only a particular subset of knowledge graph while removing all other triples.

Table \ref{table:abalation_study_1} and Table \ref{table:abalation_study_2} show the relative performance of models trained on specific subsets of the knowledge graph. It can be observed that on removing triples related to BIOKG while keeping triples related GO and PRO ontologies leads to a modest drop in performance. On the other hand training models with only the BIOKG triples without any ontology information leads to a pretty significant drop in performance. Further, training models with only GO or only PRO ontologies leads to a pretty significant increase in performance. 

These results though interesting, are not entirely suprising. When we integrated the GO and PRO ontology information into our knowledge, we included the triples denoting relations all the way up to the root node of the ontology. This allowed our model to learn a much better representation of kinases and substrates in terms of their shared evolutionary, molecular and functional ancestory. BIOKG on the other hand includes triples that denote relationships only at the leaf node without following them up the ontology tree. Thus, even though it brings a lot of information, it is only useful in conjunction with a more complete picture provided by the ontologies.

\begin{table}
  \scalebox{0.80}{
    \begin{tabular}{| p{5.2cm} | p{4.2cm} |}
       \hline
       \textit{Data} & \textit{Interaction Prediction} \\
       \hline
       BIOKG complex (removed) & \textit{\textbf{0.75}} $(\pm0.01)$  \\
       \hline
       BIOKG pathways (removed) & \textit{\underline{0.77}} $(\pm0.01)$   \\
       \hline
       BIOKG diseases (removed) & \textit{\underline{0.78}} $(\pm0.01)$  \\
       \hline
       PRO (removed) & 0.79 $(\pm0.01)$  \\
       \hline
       GO biological process (removed) & 0.79 $(\pm0.03)$  \\
       \hline
       GO molecular function (removed) & 0.79 $(\pm0.01)$  \\
       \hline
       \textit{COMPLETE KG} & \textit{0.79} $(\pm0.01)$  \\
       \hline       
    \end{tabular}
  } \caption{{(Part A) Abalation study showing relative importance when a set of triples are removed from KG}}
  \label{table:abalation_study_1}
\end{table}

\begin{table}
  \scalebox{0.80}{
    \begin{tabular}{| p{5.2cm} | p{4.2cm} |}
       \hline
       \textit{Data} & \textit{Interaction Prediction} \\
       \hline
       BIOKG complex (only) & 0.60 $(\pm0.02)$  \\
       \hline
       BIOKG Pathways (only) & 0.61 $(\pm0.02)$   \\
       \hline
       BIOKG Diseases (only) & 0.63 $(\pm0.02)$  \\
       \hline
       GO Biological Process (only) & \textit{\textbf{0.84}} $(\pm0.01)$  \\
       \hline
       GO Molecular Function (only) & \textit{\underline{0.83}} $(\pm0.01)$  \\
       \hline
       PRO (only) & \textit{\underline{0.82}} $(\pm0.01)$  \\
       \hline
       \textit{COMPLETE KG} & \textit{0.79} $(\pm0.01)$  \\
       \hline       
    \end{tabular}
  } \caption{{(Part B) Abalation study showing relative importance when including only one set of triples in KG.}}
  \label{table:abalation_study_2}
\end{table}

\paragraph{Functional enrichment analysis:}
After validating our frameworks predictive performance, we studied the highest confidence predictions for kinases with the least amount of information. We retrieved the list of understudied kinases from Illuminating the Druggable Genome project (IDG) \cite{idg}. This gave us a list of 144 potenitally understudied kinases. We further filtered these kinases to only include the kinases that have at-most two recorded interactions in the iPTMnet database. This gave us a list of 68 kinases that are potentially understudied with respect to both IDG and iPTMnet. We then used the PredKinKG framework to predict novel interactions for these kinases. We filtered the predictions to only include high confidence predictions by setting the probability score cutoff at 0.95. Below we present the functional enrichment analysis of Q02779 \textit{(MAP3K10)} using its 188 novel predicted substrates.

Since the target kinase is understudied and its biological functions is poorly understood, we hypothesized that studying the functions of the predicted substrates may provide us with clues about its biological roles. For this we perform a GO enrichment analysis using STRINGS DB \cite{strings_db}. Table \ref{table:gokinase1} provides an overview of the top five GO terms (according to strength and FDR) enriched for every GO sub-ontology. 

GO enrichment analysis of the interaction network of Q02779 \textit{(MAP3K10)} suggests that it might play an important role in the maintainence and upkeep of the cellular DNA and regulation of DNA transcription. We can observe that the GO term - \textit{GO:0090240 (Positive regulation of histone h4 acetylation)} has the highest enrichment strength. Histone-H4 is a part of the nucleosome complex which is one of the fundamental structures related to DNA organization in eukaryotes. Acetylation of Histone-H4 is associated with a relaxation of the nuclear chromatin leading to an increased transcription factor binding \cite{h4_transcription} and recruitement of protein complexes for repair of double-stranded breaks in the DNA \cite{h4_dsb}. In addition to histone acetylation, we can also observe that several GO terms related to DNA damage and repair are enriched: \textit{GO:0090400 (DNA ligation involved in DNA repair), GO:0006978 (DNA damage response, signal trans- duction by p53), GO:0042771 (Intrinsic apoptotic signaling pathway in response to dna damage by p53 class mediator)}. Analayzing the GO terms related to cellular component, it is evident that the interaction partners of Q02779 \textit{(MAP3K10)} are mostly located in the nucleus near the chromosomes thus further cementing its role in DNA repair and transcription.

\begin{table}
    \scalebox{0.65}{
      \begin{tabular}{| p{2.0cm} | p{5.3cm} | p {2.0cm} | p {1.4cm} | }
        \hline
         \multicolumn{4}{|c|}{\textbf{Biological Process}} \\
         \hline
         \textit{GO:ID} & \textit{DESCRIPTION} & \textit{STRENGTH} & \textit{FDR} \\
        \hline 
        GO:0090240 & Positive regulation of histone h4 acetylation & 1.65 & 0.005\\   
        \hline
        GO:0070601 & Centromeric sister chromatid cohesion & 1.65 & 0.005\\
        \hline  
        GO:0090400 & DNA ligation involved in DNA repair  & 1.60 & 0.00091\\   
        \hline
        GO:0006978 & DNA damage response, signal transduction by p53 class mediator resulting in transcription of p21 class mediator  & 1.45 & 0.0019\\   
        \hline
        GO:0042771 & Intrinsic apoptotic signaling pathway in response to dna damage by p53 class mediator  & 1.24 & 0.0014 \\
        \hline
        \multicolumn{4}{|c|}{\textbf{Molecular Function}} \\
        \hline
        \textit{GO:ID} & \textit{DESCRIPTION} & \textit{STRENGTH} & \textit{FDR}\\
        \hline 
        GO:0031490 & Chromatin dna binding & 1.03 & 0.0003 \\
        \hline  
        GO:0070491 & Repressing transcription factor binding & 1.00 & 0.0014 \\   
        \hline
        GO:0019901 & Protein kinase binding & 0.78 & 9.3E-15 \\   
        \hline
        GO:0051721 & DNA-binding transcription factor binding & 0.78 & 6.02E-08 \\   
        \hline
        GO:0061629 & RNA polymerase II-specific DNA-binding transcription factor binding & 0.75  & 4.32E-05 \\      
        \hline
         \multicolumn{4}{|c|}{\textbf{Cellular Component}} \\
         \hline
         \textit{GO:ID} & \textit{DESCRIPTION} & \textit{STRENGTH} & \textit{FDR}\\
        \hline 
        GO:0005719 & Lateral element & 1.48 & 0.0013 \\
        \hline  
        GO:0005721 & Pericentric heterochromatin & 1.40 & 0.0003 \\   
        \hline
        GO:0000778 & Condensed nuclear chromosome kinetochore & 1.37 & 0.0029  \\   
        \hline
        GO:0000780 & Condensed nuclear chromosome, centromeric region & 1.24 & 0.0012  \\   
        \hline
        GO:0051233 & Spindle midzone & 1.16 & 0.0025 \\   
        \hline
      \end{tabular}
    }
    \caption{\scalebox{0.7}{Enriched GO terms for Q02779 (MAP3K10) interaction partners}}
    \label{table:gokinase1}
\end{table}

\section{Discussion and Future work}
In this work we have presented our framework for learning from a heterogenous knowledge graph to predict substrates for understudied kinases. We build a kinase-substrate knowledge graph by integrating data from ontologies such as GO and PRO and existing knowledge graph such as BIOKG. We then developed a novel knowledge-graph representation learning approach to learn better representations of kinases and substrates in this knowledge graph. Unlike many existing approaches, our framework can take advantage of semantic data from existing databases to exploit the knowledge of well studied kinases to make predictions for understudied kinases. We also perform an ablation study to quantify the relative importance of various components of our knowledge graph. We found that the heirarchical information from ontologies in combination with the factual information from existing knowledge graphs contributes significantly to learning a better representation of kinases and substrates.

A significant advantage of our methods over existing methods is the simplicity of the data representation. Existing machine learning systems require complex preprocessing and data transformation before the data is used for model training. These data transformations take a lot of manual effort and also have the potential to influence the model performance if done incorrectly. In our approach the data is arranged in a very simple form containing only triples which represent a discrete fact about the real world. This enables our approach to scale as the volume of data scales without requiring significant manual effort. The TripleWalk approach is very generic and does not assume any structure of the underlying knowledge graph. Hence, it can be readily repurposed to be applied in any alternative domain such as learning a representation of a knowledge graph containing drugs and predicting novel drug-drug interactions or even a non-biomedical domain such as social network analysis. To further simplify the application of the TripleWalk algorithm on existing knowledge graphs, we have also made it available as a python package for the community \cite{TripleWalk}. 

A significant shortcoming of our approach is that it can make predictions only at kinase/substrate level and not at the kinase/phosphorylation site level. Thus, as a next step of our study we plan to extend our model to make predictions at the phosphorylation site level by integrating with the approach proposed by Deznabi et al. in their DeepKinZero model. In addition to extending to model to site level, we also plan to integrate attention mechanism in our unsupervised knowledge graph learning component to get a better insight into the factors that contribute to learning a good representation of kinases and substrates. 

Since the goal of this work was to develop a system to utilize semantic data (knowledge graph) for the purpose of predicting kinase-substrate interactions, we did not perform an in direct comparision with existing kinase-substrate interaction prediction tools but rather with existing knowledge graph learning methods. A direct comparison between our tool and other the described tools is not possible due to the differences in the format of the data used for training and testing. The data for positive samples is readily available from established databases, but data about negative samples is not readily available. Thus, every tool uses its own method to generate negative samples which further complicates the comparision. A comprehensive evaluation will require a more focused approach that uses a standardized dataset with properly specified training and testing splits and negative samples. Since developing such a dataset is a non-trivial task, we plan to perform this comparision as its own independent study.

\section*{Author Contributions}
\textbf{Sachin Gavali:} Conceptualization, Data Curation, Methodology, Software, Formal analysis, Writing - Original Draft.
\textbf{Karen E. Ross:} Data Curation, Validation, Writing - Review and Editing.
\textbf{Chuming Chen:} Data Curation, Writing - Review and Editing, Supervision.
\textbf{Julie Cowart:} Data Curation, Writing - Review and Editing.
\textbf{Cathy H. Wu:} Supervision, Project administration, Funding acquisition.

\section*{Conflicts of interest}
There are no conflicts to declare

\section*{Data availablility}
The data and code used to perform the above analysis can be found at : \url{https://github.com/udel-cbcb/ikg_v2_public.git}. The code for the triple walk algorithm can be found at \url{https://github.com/udel-cbcb/triple_walk.git}: 

\section*{Acknowledgements}
This work has been supported by the National Institutes of Health (Grant-R35GM141873) and Sigma Xi Grants in Aid of
Research (Grant-G20201001105037737). We would also like to acknowledge the Data Science Institute and Center for Bioinformatics and Computational Biology at University of Delaware for providing computing resources through the Caviness and Biomix high performance computing clusters respectively.

\bibliographystyle{unsrt}
\bibliography{references}

\begin{thebibliography}{10}

\bibitem{proteome}
Young-Ki Paik, Seul-Ki Jeong, Gilbert~S. Omenn, Mathias Uhlen, Samir Hanash,
  Sang~Yun Cho, Hyoung-Joo Lee, Keun Na, Eun-Young Choi, Fangfei Yan, Fan
  Zhang, Yue Zhang, Michael Snyder, Yong Cheng, Rui Chen, György Marko-Varga,
  Eric~W. Deutsch, Hoguen Kim, Ja-Young Kwon, Ruedi Aebersold, Amos Bairoch,
  Allen~D. Taylor, Kwang~Youl Kim, Eun-Young Lee, Denis Hochstrasser, Pierre
  Legrain, and William~S. Hancock.
\newblock The chromosome-centric human proteome project for cataloging proteins
  encoded in the genome.
\newblock {\em Nature Biotechnology}, 30(3):221–223, Mar 2012.

\bibitem{Uversky_2013}
V.N. Uversky.
\newblock {\em Posttranslational Modification}, page 425–430.
\newblock Elsevier, 2013.

\bibitem{iptmnet}
Hongzhan Huang, Cecilia~N. Arighi, Karen~E. Ross, Jia Ren, Gang Li, Sheng-Chih
  Chen, Qinghua Wang, Julie Cowart, K.~Vijay-Shanker, and Cathy~H. Wu.
\newblock iptmnet: an integrated resource for protein post-translational
  modification network discovery.
\newblock {\em Nucleic Acids Research}, 46(D1):D542–D550, Jan 2018.

\bibitem{pro_ontology}
Darren~A. Natale, Cecilia~N. Arighi, Judith~A. Blake, Jonathan Bona, Chuming
  Chen, Sheng-Chih Chen, Karen~R. Christie, Julie Cowart, Peter D’Eustachio,
  Alexander~D. Diehl, Harold~J. Drabkin, William~D. Duncan, Hongzhan Huang, Jia
  Ren, Karen Ross, Alan Ruttenberg, Veronica Shamovsky, Barry Smith, Qinghua
  Wang, Jian Zhang, Abdelrahman El-Sayed, and Cathy~H. Wu.
\newblock Protein ontology (pro): enhancing and scaling up the representation
  of protein entities.
\newblock {\em Nucleic Acids Research}, 45(D1):D339–D346, Jan 2017.

\bibitem{gene_ontology}
David~P Hill, Barry Smith, Monica~S McAndrews-Hill, and Judith~A Blake.
\newblock Gene ontology annotations: what they mean and where they come from.
\newblock {\em BMC Bioinformatics}, 9(Suppl 5):S2, 2008.

\bibitem{biokg}
Brian Walsh, Sameh~K. Mohamed, and Vít Nováček.
\newblock Biokg: A knowledge graph for relational learning on biological data.
\newblock In {\em Proceedings of the 29th ACM International Conference on
  Information \& Knowledge Management}, page 3173–3180. ACM, Oct 2020.

\bibitem{kg_embedding_lp}
Mona Alshahrani, Maha~A. Thafar, and Magbubah Essack.
\newblock Application and evaluation of knowledge graph embeddings in
  biomedical data.
\newblock {\em PeerJ Computer Science}, 7:e341, Feb 2021.

\bibitem{tensor_decomposition_1}
Yuwang Ji, Qiang Wang, uan Li, and Jie Liu.
\newblock A survey on tensor techniques and applications in machine learning.
\newblock {\em IEEE Access}, 7:162950–162990, 2019.

\bibitem{tensor_decompisition_2}
Stephan Rabanser, Oleksandr Shchur, and Stephan Günnemann.
\newblock Introduction to tensor decompositions and their applications in
  machine learning.
\newblock {\em arXiv:1711.10781 [cs, stat]}, Nov 2017.
\newblock arXiv: 1711.10781.

\bibitem{transe}
Antoine Bordes, Nicolas Usunier, Alberto Garcia-Duran, Jason Weston, and Oksana
  Yakhnenko.
\newblock Translating embeddings for modeling multi-relational data.
\newblock {\em Advances in neural information processing system}, page~9, 2013.

\bibitem{conve}
Tim Dettmers, Pasquale Minervini, Pontus Stenetorp, and Sebastian Riedel.
\newblock Convolutional 2d knowledge graph embeddings.
\newblock {\em Proceedings of the AAAI Conference on Artificial Intelligence},
  32(11):20--30, Apr 2018.

\bibitem{word2vec}
Tomas Mikolov, Kai Chen, Greg Corrado, and Jeffrey Dean.
\newblock Efficient estimation of word representations in vector space.
\newblock {\em arXiv:1301.3781 [cs]}, Sep 2013.
\newblock arXiv: 1301.3781.

\bibitem{deepwalk}
Bryan Perozzi, Rami Al-Rfou, and Steven Skiena.
\newblock Deepwalk: online learning of social representations.
\newblock In {\em Proceedings of the 20th ACM SIGKDD international conference
  on Knowledge discovery and data mining}, page 701–710. ACM, Aug 2014.

\bibitem{node2vec}
Aditya Grover and Jure Leskovec.
\newblock node2vec: Scalable feature learning for networks.
\newblock In {\em Proceedings of the 22nd ACM SIGKDD International Conference
  on Knowledge Discovery and Data Mining}, page 855–864. ACM, Aug 2016.

\bibitem{metapath2vec}
Yuxiao Dong, Nitesh~V. Chawla, and Ananthram Swami.
\newblock metapath2vec: Scalable representation learning for heterogeneous
  networks.
\newblock In {\em Proceedings of the 23rd ACM SIGKDD International Conference
  on Knowledge Discovery and Data Mining}, page 135–144. ACM, Aug 2017.

\bibitem{word_embeddings}
Felipe Almeida and Geraldo Xexéo.
\newblock Word embeddings: A survey.
\newblock {\em arXiv:1901.09069 [cs, stat]}, Jan 2019.
\newblock arXiv: 1901.09069.

\bibitem{markovian}
Richard Bellman.
\newblock A markovian decision process.
\newblock {\em Indiana University Mathematics Journal}, 6(4):679–684, 1957.

\bibitem{rf}
Leo Breiman.
\newblock Random forests.
\newblock {\em Machine Learning}, 45(1):5–32, 2001.

\bibitem{negative_sampling}
Bhushan Kotnis and Vivi Nastase.
\newblock Analysis of the impact of negative sampling on link prediction in
  knowledge graphs.
\newblock {\em arXiv:1708.06816 [cs]}, Mar 2018.
\newblock arXiv: 1708.06816.

\bibitem{negatome}
Philipp Blohm, Goar Frishman, Pawel Smialowski, Florian Goebels, Benedikt
  Wachinger, Andreas Ruepp, and Dmitrij Frishman.
\newblock Negatome 2.0: a database of non-interacting proteins derived by
  literature mining, manual annotation and protein structure analysis.
\newblock {\em Nucleic Acids Research}, 42(D1):D396–D400, Jan 2014.

\bibitem{distmult}
Bishan Yang, Wen-tau Yih, Xiaodong He, Jianfeng Gao, and Li~Deng.
\newblock Embedding entities and relations for learning and inference in
  knowledge bases.
\newblock {\em arXiv:1412.6575 [cs]}, Aug 2015.
\newblock arXiv: 1412.6575.

\bibitem{asha}
Liam Li, Kevin Jamieson, Afshin Rostamizadeh, Ekaterina Gonina, Moritz Hardt,
  Benjamin Recht, and Ameet Talwalkar.
\newblock A system for massively parallel hyperparameter tuning.
\newblock {\em arXiv:1810.05934 [cs, stat]}, Mar 2020.
\newblock arXiv: 1810.05934.

\bibitem{linkphinder}
Vít Nováček, Gavin McGauran, David Matallanas, Adrián Vallejo~Blanco, Piero
  Conca, Emir Muñoz, Luca Costabello, Kamalesh Kanakaraj, Zeeshan Nawaz, Brian
  Walsh, Sameh~K. Mohamed, Pierre-Yves Vandenbussche, Colm~J. Ryan, Walter
  Kolch, and Dirk Fey.
\newblock Accurate prediction of kinase-substrate networks using knowledge
  graphs.
\newblock {\em PLoS Computational Biology}, 16(12):e1007578, Dec 2020.

\bibitem{netphosphk}
Nikolaj Blom, Thomas Sicheritz-Pontén, Ramneek Gupta, Steen Gammeltoft, and
  Søren Brunak.
\newblock Prediction of post-translational glycosylation and phosphorylation of
  proteins from the amino acid sequence.
\newblock {\em PROTEOMICS}, 4(6):1633–1649, Jun 2004.

\bibitem{scansite}
John~C. Obenauer, Lewis~C. Cantley, and Michael~B. Yaffe.
\newblock Scansite 2.0: Proteome-wide prediction of cell signaling interactions
  using short sequence motifs.
\newblock {\em Nucleic Acids Research}, 31(13):3635–3641, Jul 2003.

\bibitem{idg}
Mar 2015.

\bibitem{strings_db}
Damian Szklarczyk, Annika~L Gable, David Lyon, Alexander Junge, Stefan Wyder,
  Jaime Huerta-Cepas, Milan Simonovic, Nadezhda~T Doncheva, John~H Morris, Peer
  Bork, Lars~J Jensen, and Christian von Mering.
\newblock String v11: protein–protein association networks with increased
  coverage, supporting functional discovery in genome-wide experimental
  datasets.
\newblock {\em Nucleic Acids Research}, 47(D1):D607–D613, Jan 2019.

\bibitem{h4_transcription}
M.~Vettese-Dadey, P.~A. Grant, T.~R. Hebbes, C.~Crane-Robinson, C.~D. Allis,
  and J.~L. Workman.
\newblock Acetylation of histone h4 plays a primary role in enhancing
  transcription factor binding to nucleosomal dna in vitro.
\newblock {\em The EMBO Journal}, 15(10):2508–2518, May 1996.

\bibitem{h4_dsb}
Surbhi Dhar, Ozge Gursoy-Yuzugullu, Ramya Parasuram, and Brendan~D. Price.
\newblock The tale of a tail: histone h4 acetylation and the repair of dna
  breaks.
\newblock {\em Philosophical Transactions of the Royal Society B: Biological
  Sciences}, 372(1731):20160284, Oct 2017.

\bibitem{TripleWalk}
Sachin Gavali.
\newblock triple-walk: A pytorch extension library to perform triple walks on
  knowledge graphs.

\end{thebibliography}
\end{document}